\documentclass[aps,prl,twocolumn,showpacs,a4paper]{revtex4}
\pdfoutput=1
\usepackage{float}
\usepackage{color}
\usepackage[utf8x]{inputenc}
\usepackage{graphicx}
\usepackage{amsmath,amssymb}
\begin{document}
\title{Creating, moving and merging Dirac points with a Fermi gas in a tunable honeycomb lattice}

\author{%
Leticia Tarruell, Daniel Greif, Thomas Uehlinger, Gregor Jotzu and Tilman
Esslinger}
\affiliation{%
Institute for Quantum Electronics, ETH Zurich, 8093 Zurich, Switzerland }


\maketitle
{\bf Dirac points lie at the heart of many fascinating
phenomena in condensed matter physics, from massless electrons in graphene
to the emergence of conducting edge states in topological insulators
\cite{CastroNetoRMP09,HasanRMP10}. At a Dirac point, two energy bands
intersect linearly and the particles behave as relativistic Dirac
fermions. In solids, the rigid structure of the material sets the mass and
velocity of the particles, as well as their interactions. A different,
highly flexible approach is to create model systems using fermionic atoms
trapped in the periodic potential of interfering laser beams, a method
which so far has only been applied to explore simple lattice structures
\cite{BlochRMP08,EsslingerARCMP10}. Here we report on the creation of
Dirac points with adjustable properties in a tunable honeycomb optical
lattice. Using momentum-resolved interband transitions, we observe a
minimum band gap inside the Brillouin zone at the position of the Dirac
points. We exploit the unique tunability of our lattice potential to
adjust the effective mass of the Dirac fermions by breaking inversion
symmetry. Moreover, changing the lattice anisotropy allows us to move the
position of the Dirac points inside the Brillouin zone. When increasing
the anisotropy beyond a critical limit, the two Dirac points merge and
annihilate each other -- a situation which has recently attracted
considerable theoretical interest \cite{HasegawaPRB06,ZhuPRL07,
WunschNJP08,MontambauxPRB09,LeePRA09}, but seems extremely challenging to
observe in solids \cite{PereiraPRB09}. We map out this topological
transition in lattice parameter space and find excellent agreement with
\emph{ab initio} calculations. Our results not only pave the way to model
materials where the topology of the band structure plays a crucial role,
but also provide an avenue to explore many-body phases resulting from the
interplay of complex lattice geometries with interactions
\cite{BalentsNature10,MengNature10}. }

\begin{figure}[b!]
\includegraphics[width=1.\columnwidth]{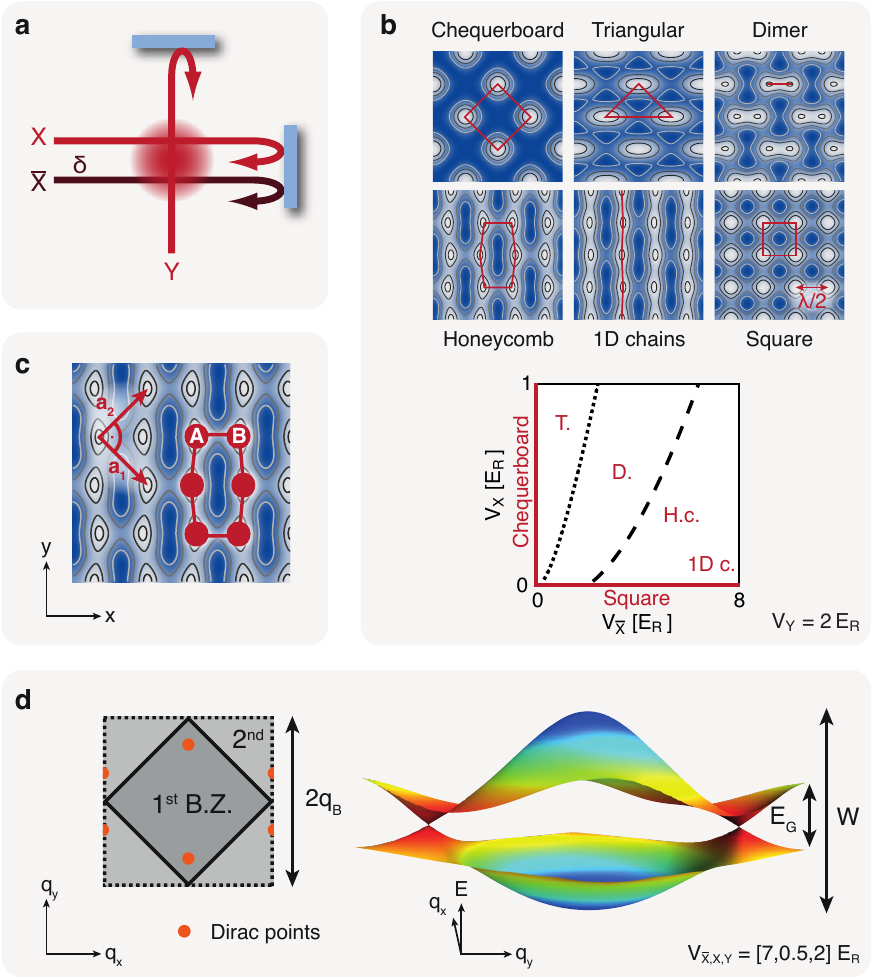}
\caption{%
{\bf Optical lattice with adjustable geometry.} {\bf a,} Three
retro-reflected laser beams of wavelength $\lambda=1064$ nm create the
two-dimensional lattice potential of equation (\ref{eqlattice}). $X$ and
$Y$ interfere and produce a chequerboard pattern, while $\overline{X}$
creates an independent standing wave. Their relative position is
controlled by the detuning $\delta$. {\bf b,} Different lattice potentials
can be realised depending on the intensities of the lattice beams, as
displayed above. The diagram below shows the accessible lattice geometries
as a function of the lattice depths $V_{\overline{X}}$ and $V_{X}$. The
transition between triangular (T.) and dimer (D.) lattices is indicated by
a dotted line. When crossing the dashed line into the honeycomb (H.c.)
regime, Dirac points appear. The limit $V_{\overline{X}}\gg V_{X,Y}$
corresponds to weakly coupled one-dimensional chains (1D c.). {\bf c,} The
real space potential of the honeycomb lattice has a 2-site unit cell
($A$,$B$ sites) and the primitive lattice vectors are perpendicular. {\bf
d,} Sketch of the first and second Brillouin zones (B.Z.) of the honeycomb
lattice, indicating the position of the Dirac points. On the right, a
three dimensional view of the energy spectrum shows the linear
intersection of the bands at the two Dirac points. We denote the full
bandwidth $W$, and the minimum energy gap at the edges of the Brillouin
zone $E_G$.}\label{fig:lattice-scheme}
\end{figure}

Ultracold Fermi gases have emerged as a versatile tool to simulate
condensed matter phenomena \cite{BlochRMP08,
GiorginiRMP08,EsslingerARCMP10}. For example, the control of interactions
in optical lattices has lead to the observation of Mott insulating phases
\cite{JoerdensNature08,SchneiderScience08}, providing new access to the
physics of strongly correlated materials. However, the topology of the
band structure is equally important for the properties of a solid. A prime
example is the honeycomb lattice of graphene, where the presence of
topological defects in momentum space -- the Dirac points -- leads to
extraordinary transport properties, even in the absence of interactions
\cite{CastroNetoRMP09}. In quantum gases, a honeycomb lattice has recently
been realised and investigated with a Bose-Einstein condensate
\cite{SoltanPanahiNaturePhys10,SoltanPanahiNaturePhys11}, but no
signatures of Dirac points were observed. Here we study an ultracold Fermi
gas of $^{40}$K atoms in a two-dimensional tunable optical lattice, which
can be continuously adjusted to create square, triangular, dimer and
honeycomb structures. In the honeycomb lattice, we identify the presence
of Dirac points in the band structure by observing a minimum band gap
inside the Brillouin zone via interband transitions. Our method is closely
related to a technique recently used with bosonic atoms to characterize
the linear crossing of two high-energy bands in a one-dimensional
bichromatic lattice \cite{SalgerPRL07}, but provides in addition momentum
resolution.

To create and manipulate Dirac points, we have developed a two-dimensional
optical lattice of adjustable geometry. It is formed by three
retro-reflected laser beams of wavelength $\lambda=1064$ nm, arranged as
depicted in Fig. 1a. The interference of two perpendicular beams $X$ and
$Y$ gives rise to a chequerboard lattice of spacing $\lambda/\sqrt{2}$. A
third beam $\overline{X}$, collinear with $X$ but detuned by a frequency
$\delta$, creates an additional standing wave of spacing $\lambda/2$. This
yields a potential of the form
\begin{eqnarray} &&V(x,y)=-V_{\overline{X}}\cos^2(k
x+\theta/2)-V_{X} \cos^2(k
x)\nonumber\\
&&-V_{Y} \cos^2(k y)-2\alpha \sqrt{V_{X}V_{Y}}\cos(k x)\cos(k
y)\cos\varphi\label{eqlattice}
\end{eqnarray}
where $V_{\overline{X}}$, $V_{X}$ and $V_{Y}$ denote the single beam
lattice depths (proportional to the laser beam intensities), $\alpha$ is
the visibility of the interference pattern and $k=2\pi/\lambda$. We can
adjust the two phases continuously, and choose $\theta=\pi$ and
$\varphi=0$ (see Methods). Varying the relative intensities of the beams
allows us to realise various lattice structures, as displayed in Fig. 1b.
In the following we focus on the honeycomb lattice, whose real space
potential is shown in Fig. 1c.

The honeycomb lattice consists of two sublattices $A$ and $B$. Therefore,
the wavefunctions are two-component spinors. Tunneling between the
sublattices leads to the formation of two energy bands, which are well
separated from the higher bands and have a conical intersection at two
quasi-momentum points in the Brillouin zone -- the Dirac points. These
points are topological defects in the band structure, with an associated
Berry phase of $\pm\pi$. This warrants their stability with respect to
lattice perturbations, so that a large range of lattice anisotropies only
changes their position inside the Brillouin zone. In contrast, breaking
the inversion symmetry of the potential by introducing an energy offset
$\Delta$ between sublattices opens an energy gap at the Dirac points,
proportional to the offset. In our implementation, the sublattice offset
$\Delta$ depends only on the value of the phase $\theta$ and can be
precisely adjusted (see Methods). As displayed in Fig. 1c and d, the
primitive lattice vectors are perpendicular, leading to a square Brillouin
zone with two Dirac points inside. Their position is symmetric around the
center \begin{figure}[p]
\includegraphics[width=1.\columnwidth]{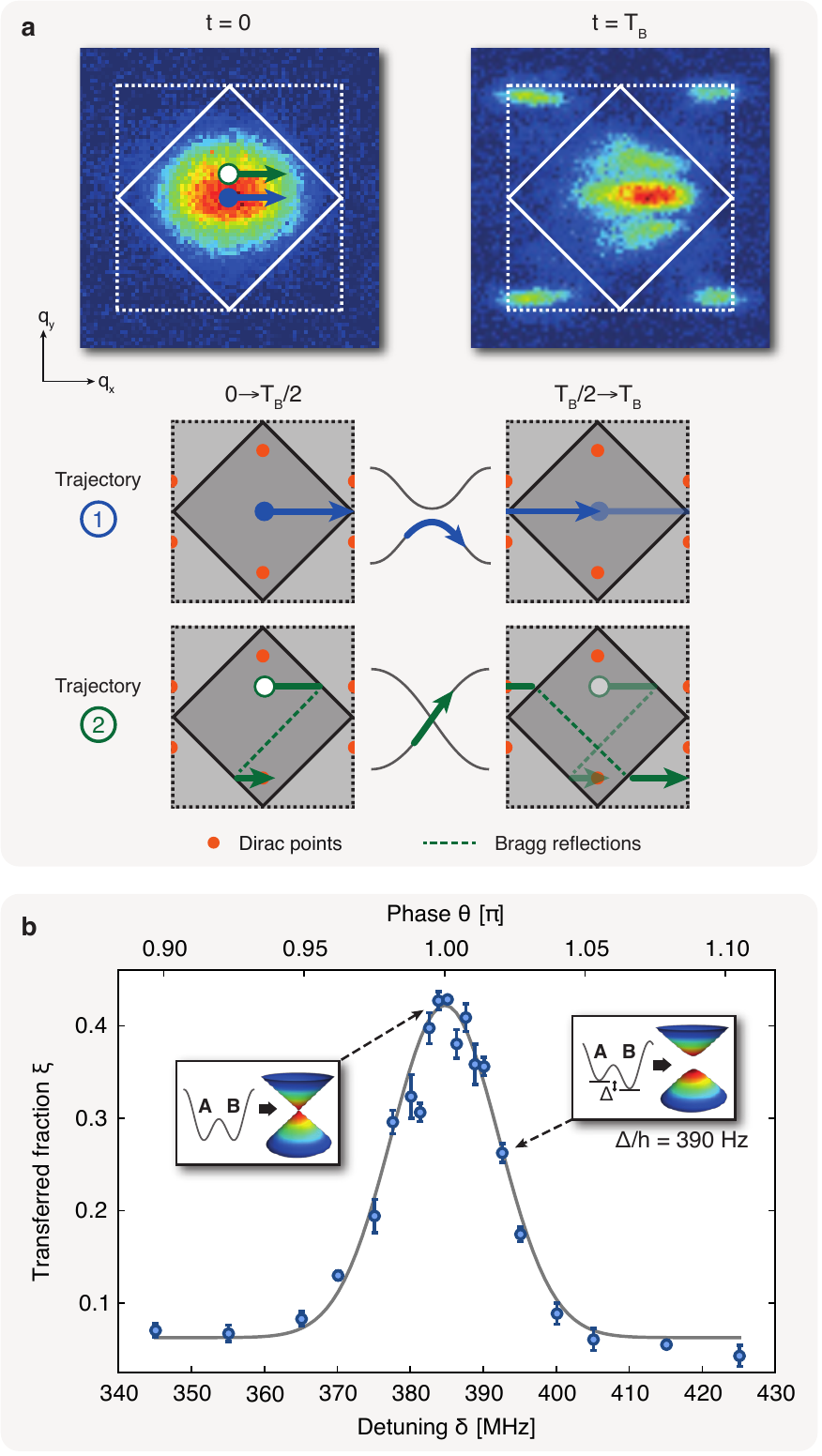}
\caption{%
{\bf Probing the Dirac points.} {\bf a,} Quasi-momentum distribution of
the atoms before and after one Bloch oscillation of period $T_B$. The
cloud explores several trajectories in quasi-momentum space
simultaneously. For trajectory 1 (blue solid circle) the atoms remain in
the first energy band. In contrast, trajectory 2 (green open circle)
passes through a Dirac point at $t=T_B/2$. There, the energy splitting
between the bands vanishes and the atoms are transferred to the second
band. When measuring the quasi-momentum distribution at $t=T_B$, these
atoms are missing in the first Brillouin zone and appear in the second
one. {\bf b,} Dependence of the total fraction of atoms transferred to the
second band $\xi$ on the detuning of the lattice beams $\delta$, which
controls the sublattice energy offset $\Delta$. The maximum indicates the
point of inversion symmetry, where $\Delta=0$ ($\theta=\pi$ in eq. (1))
and the gap at the Dirac point vanishes. Away from the peak, the atoms
behave as Dirac fermions with a tunable mass (see insets). Values and
error bars denote the mean and standard deviation of five consecutive
measurements, whereas the solid line is a Gaussian fit to the
data.}
\end{figure}
and fixed to $q_x=0$, owing to the time-reversal and reflection symmetries
of the system \cite{AsanoPRB11}.

We characterise the Dirac points by probing the energy splitting between
the two lowest energy bands through interband transitions
\cite{AndersonScience98}. The starting point of the experiment is a
non-interacting ultracold gas of $N\simeq50,000$ fermionic $^{40}$K atoms
in the $|F,m_F\rangle=|9/2,-9/2\rangle$ state. The cloud is prepared in
the lowest energy band of a honeycomb lattice with
$V_{{\overline{X}},X,Y}/E_R=[4.0(2),0.28(1),1.8(1)]$, which also causes a
weak harmonic confinement with trapping frequencies
$\omega_{x,y,z}/2\pi=[17.6(1),31.8(5),32.7(5)]$ Hz. Here $E_R=h^2 /2
m\lambda^2$, $h$ denotes the Planck constant and $m$ the mass of $^{40}$K.
By applying a weak magnetic field gradient, the atomic cloud is subjected
to a constant force $F$ along the $x$ direction, equivalent to an electric
field in solid-state systems. The atoms are hence accelerated such that
their quasi-momentum $q_x$ increases linearly up to the edge of the
Brillouin zone, where a Bragg reflection occurs. The cloud eventually
returns to the center of the band, performing one full Bloch oscillation
\cite{DahanPRL96}. We then measure the quasi-momentum distribution of the
atoms in the different bands \cite{KoehlPRL05} (see Methods).

Owing to the finite momentum width of the cloud, trajectories with
different quasi-momenta $q_y$ are simultaneously explored during the Bloch
cycle, as illustrated in Fig. 2a. For a trajectory far from the Dirac
points, the atoms remain in the lowest energy band (trajectory 1). In
contrast, when passing through a Dirac point (trajectory 2), the atoms are
transferred from the first to the second band because of the vanishing
energy splitting at the linear band crossing. When measuring the
quasi-momentum distribution, these atoms are missing in the first
Brillouin zone and appear in the second band, as can be seen in Fig. 2a.
We identify the points of maximum transfer with the Dirac points. The
energy resolution of the method is set by the characteristic energy of the
applied force \cite{DahanPRL96} $E_{B}/h= F\lambda /2h=88.6(7)$ Hz, which
is small compared to the full bandwidth $W/h=4.6$ kHz and the minimum band
gap at the edges of the Brillouin zone $E_G/h=475$ Hz.

To investigate how breaking the inversion symmetry of the lattice affects
the Dirac points, we vary the sublattice offset $\Delta$, which is
controlled by the frequency detuning $\delta$ between the lattice beams,
and measure the total fraction of atoms transferred to the second band
$\xi$. The results obtained for a honeycomb lattice with
$V_{{\overline{X}},X,Y}/E_R=[3.6(2),0.28(1),1.8(1)]$ are displayed in Fig.
2b, and show a sharp maximum in the transferred fraction. We identify this
situation as the point of inversion symmetry $\Delta=0$ ($\theta=\pi$), in
good agreement with an independent calibration (see Methods). At this
setting the band gap at the Dirac points vanishes. The population in the
second band decreases symmetrically on both sides of the peak as the gap
opens up, indicating the transition from massless to massive Dirac
fermions.

\begin{figure}[ht!]
\includegraphics[width=1.\columnwidth]{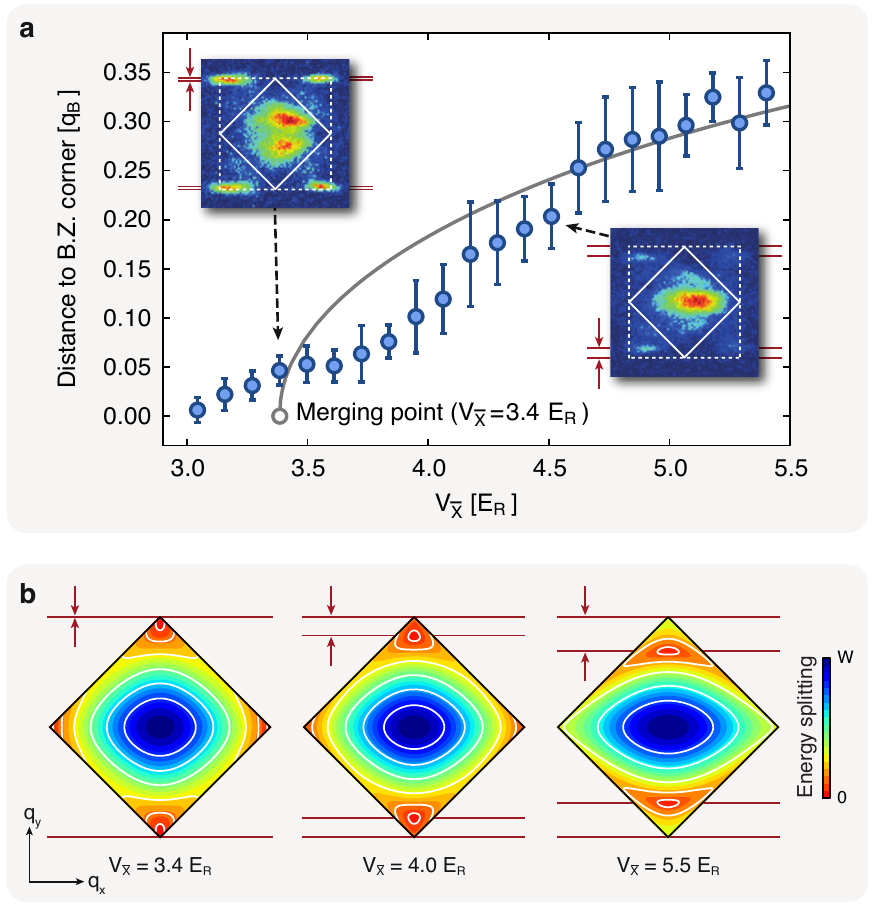}
\caption{%
{\bf Movement of the Dirac points.} {\bf a,} Distance of the Dirac points
to the corners of the Brillouin zone, as measured through
momentum-resolved interband transitions. The tunneling along the $x$
direction increases when decreasing the lattice depth $V_{\overline{X}}$.
The distance is extracted from the second band quasi-momentum distribution
after one Bloch cycle (see insets). The merging of the two Dirac points at
the corners of the Brillouin zone is signalled by a single line of missing
atoms in the first band. Values and error bars are the mean and standard
deviation of three to nine measurements. The solid line is the prediction
of a two-dimensional band structure calculation without any fitting
parameters. {\bf b,} Energy splitting between the two lowest bands. It
shows the displacement of the Dirac cones inside the Brillouin zone, as
well as their deformation depending on the lattice depth
$V_{\overline{X}}$.}\label{fig:position}
\end{figure}

The relative strength of the tunnel couplings between the different sites
of the lattice fixes the position of the Dirac points inside the Brillouin
zone, as well as the slope of the associated linear dispersion relation
\cite{HasegawaPRB06, ZhuPRL07, WunschNJP08, MontambauxPRB09,LeePRA09}.
However, the tunability of our optical lattice structure allows for
independent adjustment of the tunneling parameters along the $x$ and $y$
directions simply by controlling the intensity of the laser beams. For
isotropic tunnelings the slope of the Dirac cones is the same in all
directions, while being anisotropic otherwise. The distance of the Dirac
points to the corners of the Brillouin zone along $q_y$ can be varied
between $0$ and $q_{\mathrm{B}}/2$, whilst $q_x=0$ due to reflection
symmetry \cite{AsanoPRB11}. Here $q_{\mathrm{B}}=2 \pi/\lambda$ denotes
the Bloch wave vector.

We exploit the momentum resolution of the interband transitions to
directly observe the movement of the Dirac points. Starting from a
honeycomb lattice with $V_{\overline{X},X,Y}/E_R=[5.4(3),0.28(1),1.8(1)]$,
we gradually increase the tunneling along the $x$ direction by decreasing
the intensity of the $\overline{X}$ beam. As displayed in Fig. 3, the
position of the Dirac points continuously approaches the corners of the
Brillouin zone, as expected from an \emph{ab initio} two-dimensional band
structure calculation (see Methods).

When reaching the corners, the two Dirac points merge, annihilating each
other. There, the dispersion relation becomes quadratic along the $q_y$
axis, remaining linear along $q_x$. Beyond this critical point, a finite
band gap appears for all quasi-momenta of the Brillouin zone. This
situation signals the transition between band structures of two different
topologies, one containing two Dirac points and the other none. For
two-dimensional honeycomb lattices at half filling, it corresponds to a
Lifshitz phase transition from a semi-metallic to a band-insulating phase
\cite{ZhuPRL07, WunschNJP08}.

\begin{figure}[b!]
\includegraphics[width=1.\columnwidth]{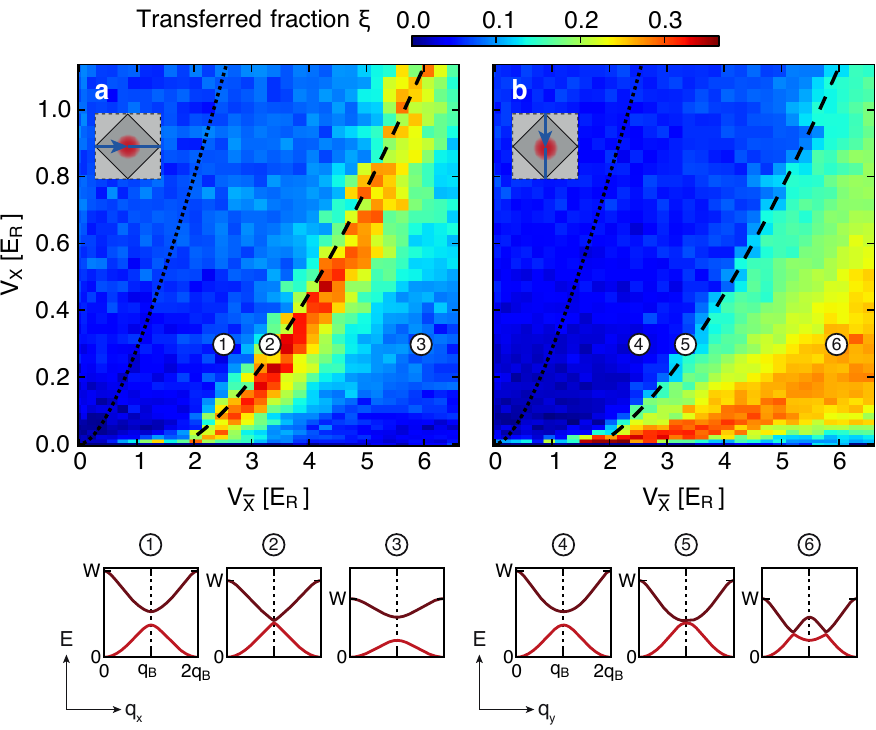}
\caption{%
{\bf Topological transition.} Fraction of atoms transferred to the second
band $\xi$ as a function of lattice depths $V_{\overline{X}}$ and $V_{X}$,
with $V_{Y}/E_R=1.8(1)$. Different lattice geometries (square,
chequerboard, triangular, dimer and honeycomb) are realised, see Fig. 1b.
We consider trajectories in quasi-momentum space along the $q_x$ and $q_y$
directions ({\bf a} and {\bf b} resp.). To maximise the transfer for the
$q_y$ trajectories, where the cloud successively passes the two Dirac
points, we set $\theta=1.013(1)\pi$. In both cases, the onset of
population transfer to the second band signals the topological transition,
where the Dirac points appear. The dashed line is the theoretical
prediction for the transition line without any fitting parameters. The
insets show cuts of the band structure along the $q_y$ axis ($q_x=0$),
illustrating the emergence of the Dirac points when increasing
$V_{\overline{X}}$.} \label{fig:phasediagram}
\end{figure}

We experimentally map out the topological transition line by recording the
fraction of atoms transferred to the second band $\xi$ as a function of
the lattice depths $V_{\overline{X}}$ and $V_{X}$, while keeping
$V_{Y}/E_R=1.8(1)$. The results are displayed in Fig. 4a. There, the onset
of population transfer to the second band signals the appearance of Dirac
points in the band structure of the lattice. The transferred fraction $\xi$
decreases for large values of $V_{\overline{X}}$, as the Dirac point
positions exceed the momentum width of the cloud.

To extend the range of our measurements and probe the Dirac points even in
this regime, we apply a force along the $y$ direction. We hence explore a
new class of trajectories in quasi-momentum space. This allows for the
investigation of very anisotropic Dirac cones, which become almost flat
along the $q_x$ direction as we approach the crossover to a
one-dimensional lattice structure ($V_{\overline{X}}\gg V_{X}$). Along the
$q_y$ trajectories the center of the cloud successively passes the two
Dirac points during the Bloch cycle, effectively realising a St\"uckelberg
interferometer \cite{KlingPRL10,ZenesiniPRA10} in a two-dimensional band
structure. As displayed in Fig. 4b, we again identify the topological
transition through the onset of population transfer to the second band.
The results for the transition line obtained for both measurement series
are in excellent agreement with \emph{ab initio} band structure
calculations.

In this work we have realised Dirac points with highly tunable properties
using ultracold fermionic atoms in a honeycomb optical lattice. A new
class of physical phenomena is now within the domain of quantum gas
experiments. Furthermore, the flexibility and novel observables of these
systems will provide new insights. For example, the unique coherence of
quantum gases offers the possibility of directly measuring the Berry phase
associated to the Dirac points by interferometric methods. Topological
order could be achieved by introducing artificial gauge fields, either via
Raman transitions \cite{LinNature09} or time-dependent lattice modulation
\cite{KitagawaPRB10}. Moreover, the exceptionally tunable lattice
potential we have developed opens up a wealth of new avenues for optical
lattice experiments. For spin mixtures with repulsive interactions, the
dynamic transition between dimer and square lattices should facilitate the
adiabatic preparation of an anti-ferromagnetic phase \cite{LubaschPRL11}
and enable the study of quantum criticality \cite{SachdevNaturePhys08}.
Additionally, the triangular and honeycomb lattices provide the
possibility to explore magnetic frustration and spin liquid phases
\cite{BalentsNature10,MengNature10}.

\section*{METHODS}

{\bf Preparation scheme.} After sympathetic cooling with $^{87}$Rb in a
magnetic trap, $2\times10^6$ fermionic $^{40}$K atoms are transferred into
a dipole trap operating at a wavelength of $826\,\text{nm}$. A balanced
spin mixture of the $|F,m_F\rangle=|9/2,-9/2\rangle$ and
$|9/2,-7/2\rangle$ states is evaporatively cooled at a magnetic field of
$197.6(1)$ G, in the vicinity of the Feshbach resonance at $202.1$ G. We
obtain typical temperatures of $0.2\,T_{\mathrm{F}}$, where
$T_{\mathrm{F}}$ is the Fermi temperature. The field is subsequently
reduced and a magnetic field gradient is used to selectively remove the
$|9/2,-7/2\rangle$ component, while levitating the $|9/2,-9/2\rangle$
atoms against gravity. This polarised Fermi gas is loaded into the
two-dimensional optical lattice within $200$ ms, and the dipole trap is
then switched off.

{\bf Tunable optical lattice.} The optical lattice is produced by the
combination of three retro-reflected laser beams of linear polarization
and wavelength $\lambda=1064$ nm, red detuned with respect to the $D_1$
and $D_2$ lines of $^{40}$K. The $\overline{X}$ and $X$ beams propagate
along the same axis, while the $Y$ beam is at $90.0(1)$. The interference
of the $X$ and $Y$ beams produces a chequerboard potential of the form
\cite{HemmerichEPL92} $V_1(x,y)=-V_{X} \cos^2(k x)-V_{Y} \cos^2(k
y)-2\alpha\sqrt{V_{X} V_{Y}}\cos(k x)\cos(k y)\cos\varphi$. The phase
between the two beams at the position of the atoms is stabilised
interferometrically to $\varphi/\pi=0.00(3)$ using a pair of additional
beams detuned from each other and from $\overline{X}, X$ and $Y$. This
results in a weak additional lattice along each axis of about $0.1E_R$.
The $\overline{X}$ beam creates a potential
$V_2(x)=-V_{\overline{X}}\cos^2(k x+\theta/2)$, where the phase $\theta$
determines the relative position of the chequerboard pattern and the
one-dimensional standing wave. We control the value of $\theta$ in the
center of the cloud by adjusting the frequency detuning $\delta$ between
the $\overline{X}$ and $X$ beams. We infer the precise value
$\theta/\delta=[\pi/384.7(6)]$ MHz${^{-1}}$ from the peak position in Fig.
2b. This is in good agreement with an independent calibration obtained
using Raman-Nath diffraction on a $^{87}$Rb Bose-Einstein condensate,
which yields $\theta/\delta=[\pi/388(4)]$ MHz${^{-1}}$. At the edges of
the cloud the phase differs by approximately $\pm10^{-4}\pi$. The total
lattice potential is given by $V_1(x,y)+V_2(x)$ and, depending on the
relative intensities of the beams, gives rise to square, chequerboard,
triangular and honeycomb lattices, as well as a staggered arrangement of
dimers \cite{Sebby-StrableyPRA06} and an array of weakly coupled
one-dimensional chains. The visibility $\alpha=0.90(5)$ and the lattice
depths $V_{\overline{X},X,Y}$ are calibrated by Raman-Nath diffraction.
The method has a systematic uncertainty of $10\%$ for the lattice depths,
whereas the statistical uncertainties are given in the main text. The
two-dimensional lattice lies in the $xy$ plane, whereas in the $z$
direction the atoms are harmonically trapped. Owing to the absence of
interactions the $z$ direction decouples. The underlying trap frequencies
in our system scale with the lattice depths according to the approximate
expressions $\omega_x\propto\sqrt{V_{Y}}$,
$\omega_y\propto\sqrt{V_{\overline{X}}+(V_{X}V_{Y}/V_{\overline{X}})}$ and
$\omega_z\propto\sqrt{V_{\overline{X}}+(V_{X}V_{Y}/V_{\overline{X}})+1.24V_{Y}}$.
For the measurements in Fig. 2a we find
$\omega_{x,y,z}/2\pi=[17.6(1),31.8(5),32.7(5)]$ Hz, as calibrated from
dipole oscillations of the cloud.

{\bf Detection.} The quasi-momentum distribution of the gas is probed
using a band-mapping technique. The optical lattice beams are linearly
ramped down in $500\,\mu$s, slowly enough for the atoms to stay
adiabatically in their band while quasi-momentum is approximately
conserved \cite{KoehlPRL05}. We then allow for $15$ ms of ballistic
expansion before taking an absorption image of the cloud.

{\bf Band structure calculations.} The energy spectrum is obtained using
an \emph{ab initio} single-particle two-dimensional numerical band
structure calculation for the homogeneous system. It therefore also takes
into account higher order tunneling terms, which are relevant for the
regime explored in this paper. In particular, they cause an asymmetry
between the two lowest bands and lead to a tilt of the Dirac cones for
certain parameter regimes.

\setlength{\parindent}{0pt}

\textbf{Acknowledgements.} We would like to thank D.~Poletti for bringing
our attention to honeycomb lattices without six-fold symmetry, and
N.~Cooper and F.~Hassler for insightful discussions. We acknowledge SNF,
NCCR-MaNEP, NCCR-QSIT, NAME-QUAM (EU, FET open), SQMS (ERC advanced grant)
and ESF (POLATOM) for funding.

\textbf{Author Contributions.} The data was taken and analysed by L.T.,
D.G., T.U. and G.J. The tunable optical lattice was built by D.G. The
experimental concept was developed by T.E. All authors contributed
extensively to the discussion of the results, as well as to the
preparation of the manuscript.

\textbf{Author Information.} Correspondence and requests for materials
should be addressed to T.E. (esslinger@phys.ethz.ch).


\begin{thebibliography}{10}
\bibitem{CastroNetoRMP09} Castro Neto, A.~H., Guinea, F., Peres, N.~M.~R., Novoselov, K.~S. \& Geim,
A.~K. The electronic properties of graphene.  \emph{Rev. Mod. Phys.} {\bf
81}, 109--162 (2009).
\bibitem{HasanRMP10} Hasan, M.~Z. \& Kane, C.~L.
Colloquium: Topological insulators. \emph{Rev. Mod. Phys.} {\bf 82},
3045--3067 (2010).
\bibitem{BlochRMP08} Bloch, I., Dalibard, J. \& Zwerger, W. Many-body physics with ultracold
gases. \emph{Rev. Mod. Phys.} {\bf 80}, 885--964 (2008).
\bibitem{EsslingerARCMP10}Esslinger, T. Fermi-Hubbard Physics with Atoms in an Optical Lattice. \emph{Annu. Rev.
Condens. Matter Phys.} {\bf 1}, 129--152 (2010).
\bibitem{HasegawaPRB06} Hasegawa, Y., Konno, R., Nakano, H. \& Kohmoto, M.
Zero modes of tight-binding electrons on the honeycomb lattice.
\emph{Phys. Rev. B} {\bf 74}, 033413 (2006).
\bibitem{ZhuPRL07} Zhu, S.-L., Wang, B. \& Duan, L.-M. Simulation and Detection of Dirac
Fermions with Cold Atoms in an Optical Lattice. \emph{Phys. Rev. Lett.}
{\bf 98}, 260402 (2007).
\bibitem{WunschNJP08} Wunsch, B. Guinea, F. \& Sols F. Dirac-point engineering and
topological phase transitions in honeycomb optical lattices. \emph{New J.
Phys.} {\bf 10}, 103027 (2008).
\bibitem{MontambauxPRB09}Montambaux, G., Piéchon, F., Fuchs, J.-N. \& Goerbig, M.~O.
Merging of Dirac points in a two-dimensional crystal. \emph{Phys. Rev. B}
{\bf 80}, 153412 (2009).
\bibitem{LeePRA09} Lee, K~L., Gr\'emaud, B., Han, R., Englert, B.-G. \& Miniatura,
C. Ultracold fermions in a graphene-type optical lattice. \emph{Phys. Rev.
A} {\bf 80}, 043411 (2009).
\bibitem{PereiraPRB09} Pereira, V.~M., Castro Neto, A.~H. \& Peres, N.~M.~R. Tight-binding
approach to uniaxial strain in graphene. \emph{Phys. Rev. B} {\bf 80},
045401 (2009).
\bibitem{BalentsNature10} Balents, L. Spin liquids in frustrated magnets. \emph{Nature} {\bf 464}, 199--208 (2010).
\bibitem{MengNature10} Meng, Z.~Y., Lang, T.~C., Wessel, S., Assaad, F.~F. \& Muramatsu, A.
Quantum spin liquid emerging in two-dimensional correlated Dirac fermions.
\emph{Nature} {\bf 464}, 847--851 (2010).
\bibitem{GiorginiRMP08} Giorgini, S., Pitaevskii, L.~P. \& Stringari, S. Theory of ultracold atomic Fermi
gases. \emph{Rev. Mod. Phys.} {\bf 80}, 1215--1274 (2008).
\bibitem{JoerdensNature08} Jördens, R., Strohmaier, N., Günter, K., Moritz, H. \&
Esslinger, T. A Mott insulator of fermionic atoms in an optical lattice.
\emph{Nature} {\bf 455}, 204--207 (2008).
\bibitem{SchneiderScience08} Schneider, U. \emph{et~al.} Metallic and Insulating Phases of Repulsively Interacting Fermions in a 3D Optical
Lattice. \emph{Science} {\bf 322}, 1520--1525 (2008).
\bibitem{SoltanPanahiNaturePhys10} Soltan-Panahi, P. \emph{et al.} Multi-component quantum gases in
spin-dependent hexagonal lattices. \emph{Nature Phys.} {\bf 7}, 434--440
(2011).
\bibitem{SoltanPanahiNaturePhys11} Soltan-Panahi, P., Lühmann, D.-S., Struck, J., Windpassinger, P. \& Sengstock, K. Quantum phase transition to unconventional
multi-orbital superfluidity in optical lattices. \emph{Nature Phys.} in
press (2011).
\bibitem{SalgerPRL07} Salger, T., Geckeler, C., Kling, S. \& Weitz,
M. Atomic Landau-Zener Tunneling in Fourier-Synthesized Optical Lattices.
\emph{Phys. Rev. Lett.} {\bf 99}, 190405 (2007).
\bibitem{AsanoPRB11} Asano K. \& Hotta, C.
Designing Dirac points in two-dimensional lattices. \emph{Phys. Rev. B}
{\bf 83}, 245125 (2011).
\bibitem{AndersonScience98} Anderson, B.~P. \& Kasevich, M.~A. Macroscopic Quantum Interference from Atomic Tunnel Arrays.
\emph{Science} {\bf 282}, 1686--1689 (1998).
\bibitem{DahanPRL96} Ben Dahan, M., Peik, E., Reichel, J., Castin, Y. \& Salomon, C.
Bloch Oscillations of Atoms in an Optical Potential. \emph{Phys. Rev.
Lett.} {\bf 76}, 4508--4511 (1996).
\bibitem{KoehlPRL05} Köhl, M., Moritz, H., Stöferle, T., Günter, K. \& Esslinger, T.
Fermionic Atoms in a Three Dimensional Optical Lattice: Observing Fermi
Surfaces, Dynamics, and Interactions. \emph{Phys. Rev. Lett.} {\bf 94},
080403 (2005).
\bibitem{KlingPRL10} Kling, S., Salger, T. , Grossert, C. \& Weitz,
M. Atomic Bloch-Zener Oscillations and Stückelberg Interferometry in
Optical Lattices. \emph{Phys. Rev. Lett.} {\bf 105}, 215301 (2010).
\bibitem{ZenesiniPRA10} Zenesini, A., Ciampini, D., Morsch, O. \&
Arimondo, E. Observation of St\"uckelberg oscillations in accelerated
optical lattices. \emph{Phys. Rev. A} {\bf 82}, 065601 (2010).
\bibitem{LinNature09} Lin, Y.-J., Compton, R.~L., Jiménez-García, K., Porto, J.~V. \&
Spielman, I.~B. Synthetic magnetic fields for ultracold neutral atoms.
\emph{Nature} {\bf 462}, 628--632 (2009).
\bibitem{KitagawaPRB10} Kitagawa, T., Berg, E., Rudner, M. \& Demler, E.
Topological characterization of periodically driven quantum systems.
\emph{Phys. Rev. B} {\bf 82}, 235114 (2010).
\bibitem{LubaschPRL11} Lubasch, M., Murg, V., Schneider, U., Cirac, J.~I. \& Ba\~nuls, M.-C.
Adiabatic Preparation of a Heisenberg Antiferromagnet Using an Optical
Superlattice. \emph{Phys. Rev. Lett.} {\bf 107}, 165301 (2011).
\bibitem{SachdevNaturePhys08} Sachdev, S. Quantum magnetism and
criticality. \emph{Nature Phys.} {\bf 4}, 173--185 (2008).
\bibitem{HemmerichEPL92} Hemmerich A., Schropp, D., Esslinger, T. \&
Hänsch, T.~W. Elastic Scattering of Rubidium Atoms by Two Crossed Standing
Waves. \emph{Europhys. Lett.} {\bf 18}, 391--395 (1992).
\bibitem{Sebby-StrableyPRA06} Sebby-Strabley, J., Anderlini, M., Jessen, P.~S. \& Porto, J.~V.
Lattice of double wells for manipulating pairs of cold atoms. \emph{Phys.
Rev. A} {\bf 73}, 033605 (2006).

\end{thebibliography}
\end{document}